\documentclass[
    ,final            
  ]
  {aipproc}

\layoutstyle{8x11single}

\newcommand{\dev}{\ensuremath{\,\mathrm{d}}}
\usepackage{amssymb}
\usepackage{amsmath}
\usepackage{epsfig}

\begin{document}

\title{From blast wave to observation}


\author{H.J. van Eerten}{
  address={Astronomical Institute 'Anton Pannekoek', Kruislaan 403, 1098SJ Amsterdam, the Netherlands}
}

\author{R.A.M.J. Wijers}{
  address={Astronomical Institute 'Anton Pannekoek', Kruislaan 403, 1098SJ Amsterdam, the Netherlands}
}

\begin{abstract}
 Gamma-ray burst (GRB) afterglows are well described by synchrotron emission originating from the interaction between a relativistic blast wave and the external medium surrounding the GRB progenitor. We introduce a code to reconstruct spectra and light curves from arbitrary fluid configurations, making it especially suited to study the effects of fluid flows beyond those that can be described using analytical approximations. As a check and first application of our code we use it to fit the scaling coefficients of theoretical models of afterglow spectra. We extend earlier results of other authors to general circumburst density profiles. We rederive the physical parameters of GRB 970508 and compare with other authors.

We also show the light curves resulting from a relativistic blast wave encountering a wind termination shock. From high resolution calculations we find that the observed transition from a stellar wind type light curve to an interstellar medium type light curve is smooth and without short-time transitory features.
\end{abstract}

\maketitle


\section{Introduction}
In the fireball model, Gamma-Ray Burst (GRB) afterglows are thought to be the result of synchrotron radiation generated by electrons during the interaction of a strongly collimated relativistic jet from a compact source with its environment (for recent reviews, see \cite{Piran2005, Meszaros2006}). Initially the resulting spectra and light curves have been modelled using only the shock front of a spherical explosion and a simple power law approximation for the synchrotron radiation (e.g. \cite{Wijers1997, Meszaros1997, Sari1998}).  One or more spectral and temporal breaks were used to connect regimes with different power law slopes. For the dynamics the self similar Blandford-McKee (BM) approximation of a relativistic explosion was used \cite{Blandford1976}. These models have been refined continuously. More details of the shock structure were included (e.g. \cite{Granot1999, Gruzinov1999}), more accurate formulae for the synchrotron radiation were used (e.g. \cite{Wijers1999}) and efforts have been made to implement collimation using various analytical approximations to the jet structure and lateral spreading behavior (see \cite{Granot2005} for an overview). On top of that, there have been studies focussing on arrival time effects (e.g. \cite{Huang2007}) and some numerical simulations (e.g. \cite{Nakar2007, Granot2001, Downes2002}).

In this paper we introduce a code to reconstruct spectra and light curves from AMRVAC, a high performance relativistic hydrodynamics code \cite{Meliani2007}. We verify our method by applying it to the analytically well understood BM solution. Because different authors have recently started using the circumstellar density structure as a fitting parameter when fitting the BM solution to afterglow data\cite{Starling2007}, we generalize existing scaling coefficient prescriptions from the literature \cite{Granot1999} from insterstellar medium (ISM, for which the inverse radial slope of the density distribution $k$ is zero) and stellar wind ($k=2$) to general $k$. \emph{These scaling coefficients are tabulated in the appendix and can be directly used when fitting to afterglow data}. We finish this part of the paper by comparing fit results to GRB 970508 using our prescription to those of other authors. 

Following this, we apply our radiation code to study the visible effect of the blast wave encountering a wind termination shock. Our simulations, done at high resolution to make sure we accurately probe the timescales at which the  encounter is expected to take place, confirm the prediction of \cite{Nakar2007} of a smooth transition between two power law regimes in the observerd light curve.

With the exception of the wind termination shock section, most of the work presented in this paper is also presented in \cite{vanEerten2009}.

\section{Description of the radiation code}
\label{peak_section}
The code takes as input a series of snapshots of relativistic hydrodynamics configurations on a grid. The grids represent a spherically symmetric fluid configuration and all grid cells are assumed to emit a fraction of their energy as radiation. This fraction of course has to be small enough not to affect the dynamics, since the post-processing approach does not allow for feedback. For the time being we restrict ourselves to the optically thin case. In this section and the next we will use BM solution for adiabatic expansion of the blast wave to provide the content of the grid snapshots. The BM solution takes two input parameters: $E_{52}$, the explosion energy in units of $10^{52}$ erg and $n_0$, the circumburst number density at a characteristic distance of $10^{17}$ cm.

Four ignorance parameters are provided to the code at runtime: $p$, $\xi_N$, $\epsilon_E$ and $\epsilon_B$, denoting respectively the slope of the relativistic particle distribution, the fraction of particles accelerated to this relativistic distribution at any given time, the fraction of thermal energy that is carried by the relativistic electrons and the fraction of thermal energy that resides in the (tangled-up) magnetic field. 

In this work we consider synchrotron radiation only. All grid cells contain a macroscopic number of radiating particles and the radiation from these particle distributions is calculated following \cite{Sari1998} and \cite{ Rybicki}, but with two important differences: the transition to the lab frame is postponed as long as possible and no assumption about the dynamics of the system is used anywhere as this should be provided by the snapshot files. 

For the emitted power per unit frequency of a typical electron we have
\begin{equation}
 \frac{ \dev P'_{<e>}}{\dev \nu'} (\nu') = \frac{p-1}{2} \cdot \frac{\sqrt{3}{q_e}^3B'}{m_ec^2} \cdot 
    \mathcal{Q} \left( \frac{\nu'}{\nu'_{cr,M}}, \frac{\nu'}{\nu'_{cr,m}}\right),
\label{ensemble_electronpower_equation}
\end{equation}

Here $q_e$ denotes the electron charge, $m_e$ the electron mass and $B'$ the local magnetic field strength. Comoving quantities are primed. The function $\mathcal{Q}$ contains the \emph{local} shape of the spectrum, including the local effect of electron cooling, resulting in $\mathcal{Q} \propto \nu^{1/3}$ for values below $\nu_{cr,m}$, in $\mathcal{Q} \propto \nu^{(1-p)/2}$ between $\nu_{cr,m}$ and $\nu_{cr,M}$ and an exponential drop beyond $\nu_{cr,M}$. It incorporates an integration over all pitch angles between electron velocities and the local magnetic field and an integration over the accelerated particle distribution. We use a power law particle distribution with a lower cut-off Lorentz factor $\gamma_m$ and an upper cut-off Lorentz factor $\gamma_M$. Directly behind the shock front the lower cut-off is determined by the downstream density, thermal energy and the ignorance parameters, while the upper cut-off is initially set to infinity. The evolution of the accelerated particle distribution when a shocked fluid parcel moves further downstream is determined by adiabatic cooling and synchrotron radiation losses. The characteristic frequencies $\nu'_{cr,M}$ and $\nu'_{cr,m}$ are related to the bounding Lorentz factors $\gamma_M$ and $\gamma_m$ via $\nu' \propto B \cdot \gamma^2$. More details on equation (\ref{ensemble_electronpower_equation}), the critical frequencies and the full shape of $Q$ can be found in \cite{vanEerten2009}.

We emphasize that $\mathcal{Q}$ describes the \emph{local} synchrotron spectrum for a location where all electrons have undergone exactly the same amount of cooling. The observed spectrum from the entire blast wave is a superposition of many such spectra. The result of this superposition is that the exponential drop beyond the locally differing $\nu_{cr,M}$ gets smoothened into a steepening of the power law slope of the spectrum by a factor $-1/2$, and the different $\nu_{cr,M}$ values together determine the position of the \emph{cooling break}, beyond which this steepening of the slope sets in.

Assuming isotropic radiation in the comoving frame, we arrive at
\begin{equation}
\frac{ \dev^2 P_{<e>}' } {\dev \nu' \dev \Omega' } ( \nu' ) = \frac{1}{4 \pi} \frac{ \dev P'_{<e>}  }{\dev \nu'}(\nu')
\end{equation}
per solid angle $\Omega'$.

To get to the \emph{received} power per unit volume in the lab frame, we have to apply the correct beaming factors, Doppler shift the frequency and multiply the above result for a single particle with the lab frame particle density:
\begin{equation}
 \frac{ \dev^2 P_V }{\dev \nu \dev \Omega}( \nu' (\nu) ) = \frac{\xi_N n}{\gamma^3 (1-\beta \mu)^3} \cdot \frac{ \dev^2 P_{<e>}' } {\dev \nu' \dev \Omega' } ( \nu \gamma (1 - \beta \mu)  ),
\end{equation}
with $\mu$ now denoting the cosine of the angle between the fluid velocity and the observer (unprimed, so measured in the lab frame), $\beta$ the fluid velocity in units of $c$ and $n$ the number density.

Finally, the flux the observer receives at a given observer time is given by
\begin{equation}
 F(\nu) = \frac{1}{r_{obs}^2} \int \frac{ \dev^2 P_V }{\dev \nu \dev \Omega}( \nu' (\nu) ) ( 1 - \beta \mu ) c \dev A \dev t_e.
\label{integral}
\end{equation}
Here $r_{obs}$ is the observer distance\footnote{For cosmological distances $r_{obs}$ denotes the luminosity distance and redshift terms $(1+z)$ need to be inserted in the appropiate places in the equations.}, approximately the same for all fluid cells (though the differences in arrival times \emph{are} taken into account). The area $A$ denotes the \emph{equidistant surface}. For every emitting time $t_e$ a specific intersecting (with the radiating volume) surface exists from which radiation arrives exactly at $t_{obs}$. The integration over the emission times $t_e$ (represented in the different snapshot files) requires an extra beaming factor and a factor of $c$ to transform the total integral to a volume integral.

\section{scaling coefficients and application to GRB970508}

Especially for high Lorentz factors, the shape of the spectrum is dominated by the radiation coming from a very thin slab right behind the shock front. The observed emission from this slab depends on the various model parameters via power laws and a heuristic fit function can be constructed with the gradual transition between regimes being handled by a free parameter, the sharpness factor $s$. While the proportionalities of this function can be determined via simple scaling arguments, we have to set its scale using our radiation code. Also, in more detailed calculations like those done here the gradual transitions are included automatically and we can use this to provide the correct dependence of $s$ on $p$ and $k$. This eliminates $s$  as a free parameter, simplifying the fit to the data and allowing the shape of the transition to help determine whether a particular model fits the data or not. Depending on the order of the peak frequency $\nu_m$ and $\nu_m$ we have two options for a fit function.

If the cooling break lies beyond the peak frequency (`spectrum 1', for easy comparison to \cite{Granot1999}), the flux is best approximated (i.e. valid up to a few percent) by
\begin{equation}
F(\nu) = F_{m,1} \cdot \left[ \left( \frac{\nu}{\nu_{m,1}} \right)^{\textstyle -\frac{s_{m,1}}{3}} + \left( \frac{\nu}{\nu_{m,1}} \right)^{\textstyle -\frac{s_{m,1}(1-p)}{2}} \right]^{\textstyle -\frac{1}{s_{m,1}}} 
 \cdot \left[ 1 + \left( \frac{\nu}{\nu_{c,1}} \right)^{s_{c,1}/2} \right]^{\textstyle -\frac{1}{s_{c,1}}}.
\label{F1smooth_cooling_equation}
\end{equation}
Here $F_{m,1}$ is the flux at the synchrotron peak frequency for an infinite sharp break (i.e. the point where the two power laws describing region $D$ and $G$ would intersect when extrapolated), $\nu_{m,1}$ the synchrotron peak frequency and $\nu_{c,1}$ the cooling frequency. The peak flux and critical frequencies depend on the physical input for the BM solution (explosion energy and circumburst density profile) and the ignorance parameters. These dependencies are summarized in the appendix. The sharpness functions $s_{m,1}$ and $s_{c,1}$ are given by
\begin{equation}
 s_{m,1} = 2.2 - 0.52 p, \qquad
s_{c,1} = 1.6 - 0.38 p -0.16 k + 0.078 pk. 
\end{equation}

When the order of the breaks is reversed (`spectrum 5') the smooth power law for both breaks is given by
\begin{equation}
F(\nu) = F_{c,5} \cdot \left[ \left( \frac{\nu}{\nu_{c,5}} \right)^{\textstyle -\frac{s_{c,5}}{3}} + \left( \frac{\nu}{\nu_{c,5}} \right)^{\textstyle \frac{s_{c,5}}{2}} \right]^{\textstyle -\frac{1}{s_{c,5}}} 
 \cdot \left[ 1 + \left( \frac{\nu}{\nu_{m,5}} \right)^{\textstyle s_{m,5}\cdot \frac{p-1}{2}} \right]^{\textstyle -\frac{1}{s_{m,5}}}.
\end{equation}
where $F_{c,5}$ denotes the peak flux for infinite sharpness $s_{c,5}$ and the prescriptions for the sharpness are
\begin{equation}
 s_{c,5} = 0.66 - 0.16k, \qquad
 s_{m,5} = 3.7 - 0.94p + 3.64k - 1.16pk.
\end{equation}
Once again these are valid up to a few percent. Given their accuracies, all sharpness prescriptions are consistent with \cite{Granot2002}.

Various authors have used flux scaling equations to derive the physical properties of GRB 970508 from afterglow data \cite{Galama1999, Granot2002, Yost2003, vanderHorst2008}. This provides us with a context to illustrate the scaling laws presented in this paper. We will use the fit parameters obtained from broadband modeling by \cite{vanderHorst2008}. They have fit simultaneously in time and frequency while keeping $k$ as a fitting parameter. Because the only model dependencies that have been introduced by this approach are the scalings of $t$ and $\nu$ (and no scaling coefficients), their fit results are still fully consistent with our flux equations. Using the cosmology $\Omega_M = 0.27$, $\Omega_\Lambda=0.73$ and Hubble parameter $H_0=71$ km s$^{-1}$ Mpc$^{-1}$, they have for the observer distance in units of $10^{28}$ cm $r_{obs,28} = 1.635$ and redshift $z =0.835$ \cite{Metzger1997}, leading, at $t_{obs,d} = 23.3$ days, to $\nu_{c,1} = 9.21\cdot10^{13}$ Hz, $\nu_{m,1} = 4.26\cdot10^{10}$ Hz, $F_{m,1} = 0.756$ mJy, $p = 2.22$ and $k = 0.0307$. 

Both \cite{vanderHorst2008} and \cite{Galama1999} take for the hydrogen mass fraction of the circumburst medium $X=0.7$, which in our flux equations is mathematically equivalent (though conceptually different) to setting $\xi_N = (1+X)/2 = 0.85$. Unfortunately this still leaves us with four variables to determine ($\epsilon_B$, $\epsilon_E$, $E_{52}$, $n_0$) and only three constraints (peak flux, cooling and peak frequency). From a theoretical study of the microstructure of collisionless shocks \cite{Medvedev2006} arrives at $\epsilon_E \backsim \sqrt{\epsilon_B}$ and we include this as an additional constraint to obtain a closed set of equations.

For the values quoted above we obtain: $E_{52} = 0.155$, $n_0 = 1.28$, $\epsilon_B = 0.1057$, $\epsilon_E = 0.325$. For comparison we give some of the values obtained by other authors. \cite{Galama1999} obtain for the ISM case: $E_{52} = 3.5$, $n_0 = 0.03$, $\epsilon_B = 0.09$, $\epsilon_E = 0.12$. \cite{Granot2002} obtain for the ISM case: $E_{52} = 0.12$, $n_0 = 22$, $\epsilon_B = 0.012$, $\epsilon_E = 0.57$. Both use $p=2.2$. Finally \cite{vanderHorst2008} obtain for $k=0.0307$: $E_{52} = 0.435$, $n_0 = 0.0057$, $\epsilon_B = 0.103$, $\epsilon_E = 0.105$.  This comparison serves to emphasize the sensitivity of the inferred blast waves physics on the scaling coefficients of the model.

\section{Wind termination shock encounter}
\begin{figure}[h]
\includegraphics[angle=0, width=0.49\textwidth]{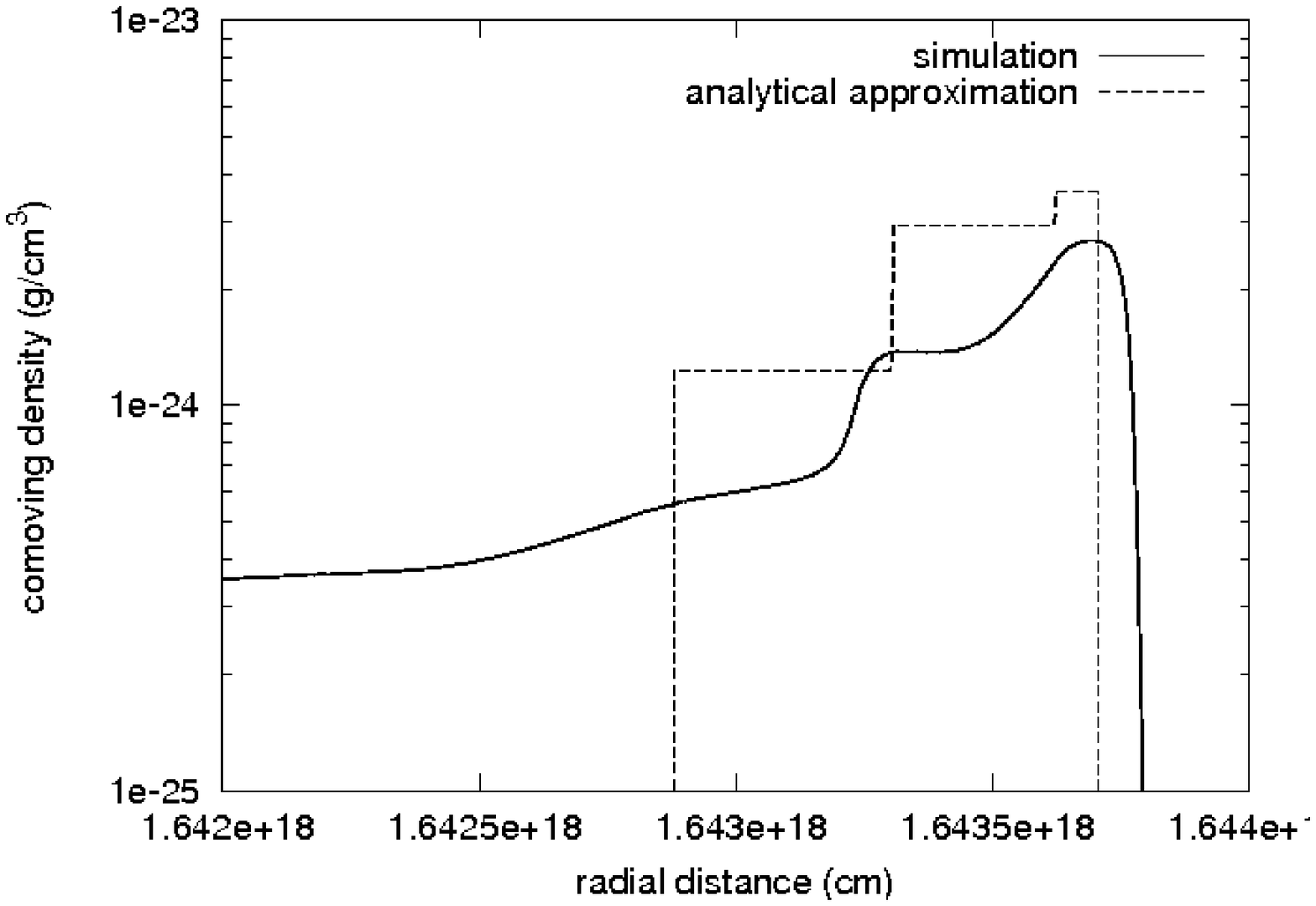}
\hspace{10mm}
\includegraphics[angle=0, width=0.49\textwidth]{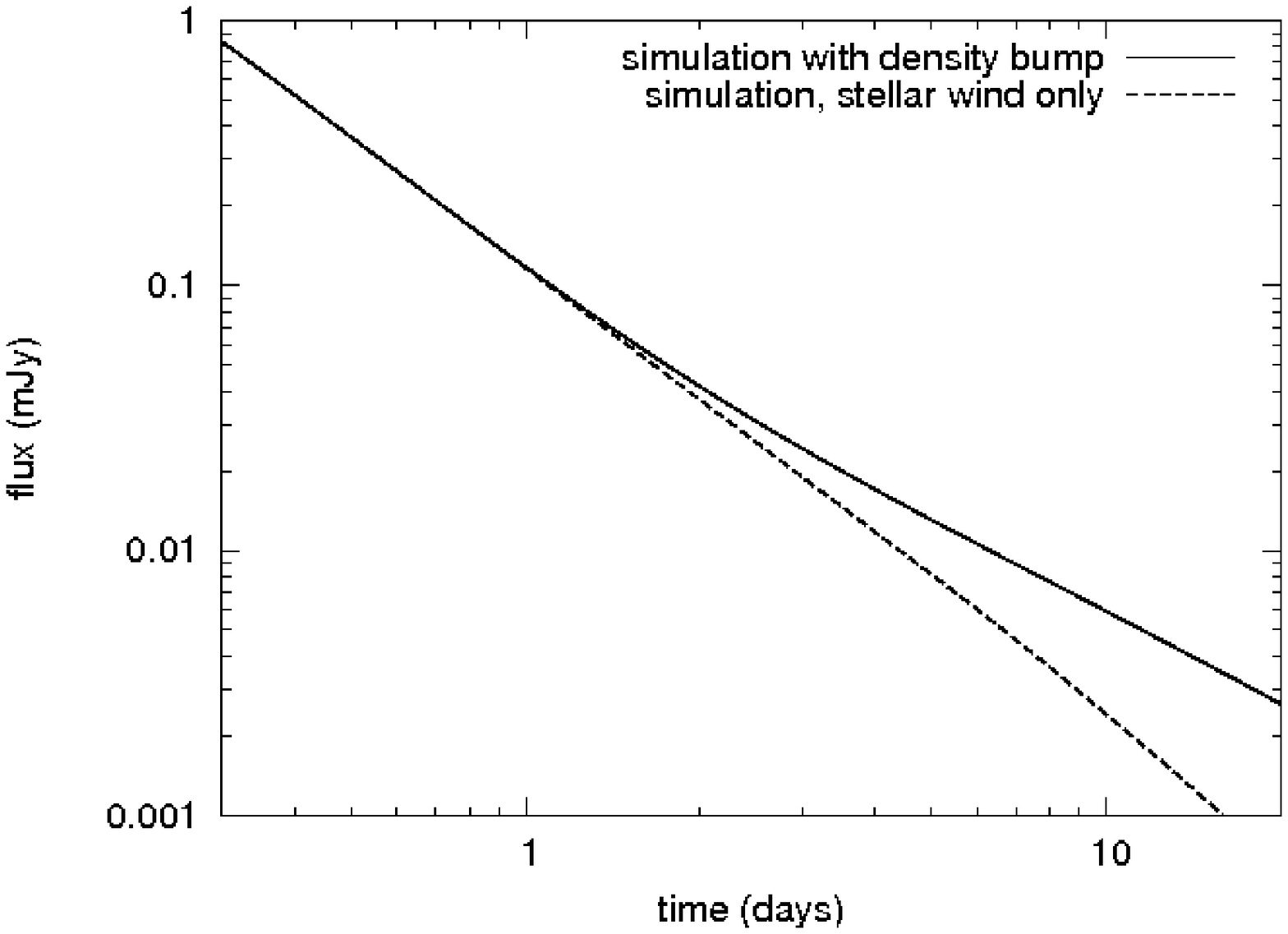}
\caption{Results for wind termination shock encounter. The left figure shows a comoving density snapshot of the shock profile during the encounter at emission time $t_e = 5.49 \cdot 10^7$ s. The schematic description of \cite{Peer2006} is shown for comparison and a clear difference between densities further downstream is visible. It should be noted that all different areas of the shock are resolved. For example, the forward shock region (smallest, rightmost region) is resolved by $\backsim 30$ cells. The right figure shows the resulting light curve at $5 \cdot 10^{14}$ Hz, with 100 data points devoted to 0.3 - 1 day and 100 datapoints to the following 19 days. A smooth transition to the power law behaviour corresponding to a BM shock wave expanding into a homogeneous environment is visible. To get complete coverage of all observer times, radiation calculated directly from the BM solution for the blast wave at Lorentz 200 down to 23 has been added to the observed flux.}
\label{termination_shock_figure}
\end{figure}

Different predictions exist in the literature concerning the observable effect of a relativistic blast wave encountering a wind-termination shock. A short transitory feature is predicted by \cite{Peer2006}, whereas \cite{Nakar2007} predict a smooth transition. \cite{Peer2006} present a detailed analysis of the reverse shock behaviour during the encounter, whereas \cite{Nakar2007} combine an analysis in a planar geometry with simulation results. We perform an RHD  simulation using AMRVAC and the initial conditions suggested by \cite{Peer2006}, with enough temporal and spatial resolution to resolve any short transity features occuring on the time scales predicted by \cite{Peer2006} (i.e. a few hours in observer time). The output of this simulation is then studied using the radiation code presented in this paper and \cite{vanEerten2009}, for an observer frequency of $5 \cdot 10^{14}$ Hz (optical).

The resulting light curve shows \emph{no short transitory feature} but a smooth transition instead, confirming lower resolution results from \cite{Nakar2007}, as can be seen from fig. \ref{termination_shock_figure}. The discrepancy between our results and \cite{Peer2006} can be attributed partly to the differences in the downstream fluid density profiles between the two studies as illustrated in fig. \ref{termination_shock_figure}. A more detailed argument will be presented elsewhere \cite{vanEerten2009b}. 

Following is a short summary of the initial conditions of the simulation. We start with a BM blastwave profile at shock Lorentz factor 23, for a shock with $E_{52} = 10.0$, $n_0 = 3$, $k = 2$. We position the wind termination shock at a radial distance of $1.6 \cdot 10^{18}$ cm. At this point the density increase by a factor 4 and remains fixed further outward. The ignorance parameters used in the radiation calculation are $p = 2.5$, $\xi_N = 1.0$, $\epsilon_E = 0.1$ and $\epsilon_B = 0.01$. We have ignored electron cooling (i.e. kept $\gamma_M$ at infinity) and differ from \cite{Peer2006} in that we do not take the slight increase in the magnetic field in the reverse shock region into account. This increase (approximately a factor $1.2^p$) is not sufficient to explain the reported transitory feature and its omission does not alter our conclusions.

The grid resolution is determined by the number of base cells (120) and the maximum refinement levels (i.e. the number of times a cell can be split in two to increase resolution, 15 in our simulation). This grid represents a radial size of $6 \cdot 10^{18}$ cm, and therefore the pre-break shock width ($\Delta R$ at Lorentz factor 23) is resolved at an effective resolution of 1200 cells. The temporal resolution is $1.556 \cdot 10^3$ s (the encounter lasts $\backsim 1.4 \cdot 10^7$ s in emission time).


\begin{theacknowledgments}
 This research was supported by NWO Vici grant 639.043.302 (RAMJW) and NOVA project 10.3.2.02 (HJvE). We gratefully acknowledge assistance from Zakaria Meliani with AMRVAC and thank Asaf Pe'er for feedback and discussion.
\end{theacknowledgments}



\bibliographystyle{aipproc}   

\begin{thebibliography}{9}

\bibitem{Blandford1976} R.D. Blandford and C.F. McKee. 1976, Phys. Fluids 19, 8
\bibitem{Downes2002} T.P. Downes, P. Duffy and S.S. Komissarov. 2002, MNRAS 332, 144-154
\bibitem{vanEerten2009} H.J. van Eerten, R.A.M.J. Wijers. 2009. MNRAS accepted. ArXiv: 0810.2250
\bibitem{vanEerten2009b} H.J. van Eerten, Z. Meliani, 2009. To be submitted,
\bibitem{Galama1999} T.J. Galama, R.A.M.J. Wijers, P.M. Vreeswijk et al. 1999, A\&A Suppl. S. 138, 451
\bibitem{Granot1999} J. Granot, T. Piran and R. Sari. 1999, ApJ, 513, 679 (astro-ph/9806192)
\bibitem{Granot2001} J. Granot et al. 2001, Gamma-Ray Bursts in the Afterglow Era: Proceedings of the International Workshop Held in Rome, Italy, 17-20 October 2000, ESO ASTROPHYSICS SYMPOSIA. ISBN 3-540-42771-6. Edited by E. Costa, F. Frontera, and J. Hjorth. Springer-Verlag, 2001, p. 312 (astro-ph/0103038)
\bibitem{Granot2002} J. Granot and R. Sari. 2002, ApJ, 568, 820 (astro-ph/0108027)
\bibitem{Granot2005} J. Granot. 2005, Rev. Mex. A\&A accepted (astro-ph/0610379)
\bibitem{Gruzinov1999} A. Gruzinov and E. Waxman. 1999, ApJ, 511, 852 (astro-ph/9807111)
\bibitem{vanderHorst2008} A.J. van der Horst. R.A.M.J. Wijers and L. van den Horn. A \& A submitted (2007)
\bibitem{Huang2007} Y. Huang, Y. Lu, A.Y.L. Wong and K.S. Cheng. 2007, ChJAA, 7(3),397 (astro-ph/0701846)
\bibitem{Medvedev2006} M. Medvedev. 2006, ApJ, 651, L9 (astro-ph/0609347)
\bibitem{Meliani2007-2} Z. Meliani. R. Keppens. 2007, A \& A 467, L41 (arXiv:0704.2461)
\bibitem{Meliani2007} Z. Meliani, R. Keppens, F. Casse, D. Giannios. 2007, MNRAS, 376(3), 1189 (astro-ph/0701434)
\bibitem{Metzger1997}M.R. Metzger, S.G. Djorgovski, S.H. Kulkarni et al. 1997, Nature, 387, 878
\bibitem{Meszaros1997} P. M\'esz\'aros and M. Rees. 1997, ApJ, 476, 232 (astro-ph/9606043)
\bibitem{Meszaros2006} P. M\'esz\'aros. 2006, Rept. Prog. Phys., 69, 2259 (astro-ph/0605208)
\bibitem{Nakar2007} E. Nakar and J. Granot. 2007, MNRAS, 380, 1744 (astro-ph/0606011)
\bibitem{Peer2005} A. Pe'er and E. Waxman. 2005, ApJ, 628, 857 (astro-ph/0409539)
\bibitem{Peer2006} A. Pe'er and R.A.M.J. Wijers. 2006, ApJ, 643, 1036 (astro-ph/0511508) 
\bibitem{Piran2005} T. Piran. 2005, Rev. Mod. Phys., 76, 1143 (astro-ph/0405503)
\bibitem{Rybicki} G.B. Rybicki, A.P. Lightman, \emph{Radiative Processes in Astrophysics} Wiley \& Sons, 1979
\bibitem{Sari1998} R. Sari, T. Piran, R. Narayan. 1998, ApJ, 497, L17 (astro-ph/9712005)
\bibitem{Starling2007} R.L.C. Starling, A.J. van der Horst, E. Rol, R.A.M.J. Wijers, C. Kouveliotou, K. Wiersema, P.A. Curran and P. Weltevrede. 2007, ApJ accepted (arXiv:0704.3718)
\bibitem{Wijers1999} R.A.M.J. Wijers and T.J. Galama. 1999, ApJ, 523, 177 (astro-ph/9805341)
\bibitem{Wijers1997} R.A.M.J. Wijers, M. Rees, P. M\'esz\'aros. 1997, MNRAS, 288, L51 (astro-ph/9704153)
\bibitem{Yost2003} S.A. Yost, F.A. Harrison, R. Sari and D.A. Frail. 2003, ApJ, 597: 459 (astro-ph/0307056)





\end{thebibliography}


\appendix
\section{APPENDIX: scaling coefficients}
\begin{table}
 \caption{Constants setting scale of flux}
  \begin{tabular}{|r|l|l|l|l|l|}
    \hline
       & D & G & H & F & E \\
    \hline
    0 & $5.12\cdot10^{-17}$ & $2.78 \cdot 10^{-31}$ & $5.68 \cdot 10^{-1}$ & $1.16 \cdot 10^{30}$ & $2.95 \cdot 10^{-16}$ \\
    $k$ & $1.18 \cdot 10^4$ & $4.54 \cdot 10^{7}$ & $6.94 \cdot 10^{-1}$ & $1.36 \cdot 10^{-8}$ & $2.04 \cdot 10^4$ \\
    $kk$ & $9.01 \cdot 10^{-1}$ & $8.95 \cdot 10^{-1}$ & $9.27 \cdot 10^{-1}$ & $1.01$ & $9.41 \cdot 10^{-1}$ \\
    $p$ & & $2.25 \cdot 10^{32}$ & $5.40 \cdot 10^{30}$ & & \\
    $pk$ & & $7.27 \cdot 10^{-9}$ & $1.65 \cdot 10^{-8}$ & & \\
    $pkk$ & & $9.41 \cdot 10^{-1}$ & $1.06$ & & \\
    $pp$ & & $1.77$ & $2.99$ & & \\
    $ppk$ & & $8.07 \cdot 10^{-1}$ & $7.01 \cdot 10^{-1}$ & & \\
    $ppkk$ & & $1.03$ & $1.01$ & & \\
   \hline
  \end{tabular}
\label{coefficient_values_table}
\end{table}
Below we provide the scaling coefficients and model fit functions for general $k$ values. The energy $E_{52}$ has been normalized to units of $10^{52}$ erg, the luminosity distance $r_{obs,28}$ to $10^{28}$ cm and the observer time $t_{obs,d}$ is expressed in days. The leftmost peak fluxes for infinitely sharp transitions are:

\begin{eqnarray}
  F_{m,1} &  = & C_D \cdot \left( \frac{ C_G}{C_D} \right)^{\frac{2}{3p-1}} \cdot \frac{ \xi_N }{r_{obs,28}^2} \cdot \epsilon_B^{\frac{1}{2}} \cdot n_0^{\frac{2}{4-k}} \cdot E_{52}^{\frac{8-3k}{2(4-k)}} \cdot t_{obs,d}^{\frac{-k}{2(4-k)}} \cdot (1+z)^{\frac{8-k}{2(4-k)}} \textrm{ mJy,} \\
 F_{c,5} & = & C_E \cdot \left( \frac{ C_F}{C_E} \right)^{\frac{2}{5}} \cdot \frac{\xi_N}{r_{obs,28}^2} \cdot \epsilon_B^{\frac{1}{2}} \cdot n_0^{\frac{2}{4-k}} \cdot E_{52}^{\frac{8-3k}{2(4-k)}} \cdot t_{obs,d}^{\frac{-k}{2(4-k)}} \cdot (1+z)^{\frac{8-k}{2(4-k)}} \textrm{ mJy.} \\
\end{eqnarray}
The critical frequencies for the different regimes are given by:
\begin{eqnarray}
 \nu_{m,1} & = & \left( \frac{C_G}{C_D} \right)^{6/(3p-1)} \cdot \left( \frac{\epsilon_E}{\xi_N} \right)^2 \cdot \epsilon_B^{1/2} \cdot E_{52}^{1/2} \cdot t_{obs,d}^{-3/2} \cdot (1+z)^{1/2} \textrm{ Hz,} \\
 \nu_{c,1} & = & \left( \frac{ C_H }{C_G} \right)^2 \cdot\epsilon_B^{-3/2} \cdot n_0^{\frac{-4}{4-k}} \cdot E_{52}^{\frac{3k-4}{2(4-k)}} \cdot t_{obs,d}^{\frac{-4+3k}{2(4-k)}} \cdot (1+z)^{-\frac{4+k}{2(4-k)}} \textrm{ Hz,} \\
 \nu_{c,5} & = & \left( \frac{C_F}{C_E} \right)^{6/5} \cdot \epsilon_B^{-3/2} \cdot n_0^{-4/(4-k)} \cdot E_{52}^{\frac{3k-4}{2(4-k)}} \cdot t_{obs,d}^{\frac{-4+3k}{2(4-k)}} \cdot (1+z)^{-\frac{4+k}{2(4-k)}} \textrm{ Hz,} \\
 \nu_{m,5} & = & \left( \frac{C_H}{C_F} \right)^{2/(p-1)} \cdot \left( \frac{ \epsilon_E }{\xi_N} \right)^2 \cdot \epsilon_B^{1/2} E_{52}^{1/2} \cdot t_{obs,d}^{-3/2} \cdot (1+z)^{1/2} \textrm{ Hz.} 
\end{eqnarray}
The flux functions and critical frequencies still contain the coefficients $C_D$, etc. They are listed below, with  $C_{D0}$ etc. denoting purely numerical constants, whose values are tabulated in table \ref{coefficient_values_table}.

\begin{eqnarray}
 C_D & \equiv & \frac{(p-1)}{3p-1} \cdot \left( C_{D0} C_{Dk}^k C_{Dkk}^{k^2} \right)^{1/(4-k)} \cdot \frac{1}{3-k} \cdot \left( \frac{p-2}{p-1} \right)^{-2/3} \cdot (17-4k)^{\frac{10-4k}{3(4-k)}} \cdot (4-k)^{\frac{2-k}{4-k}} \\
 C_E & \equiv & \left( C_{E0} C_{Ek}^k C_{Ekk}^{k^2} \right)^{1/(4-k)} \cdot \frac{1}{3-k} \cdot (17-4k)^{\frac{-6k+14}{3(4-k)}} \cdot (4-k)^{\frac{2-3k}{3(4-k)}} \\
 C_F & \equiv &  \cdot \left( C_{F0} C_{Fk}^k C_{Fkk}^{k^2} \right)^{1/(4-k)} \cdot \frac{1}{3-k} \cdot (17-4k)^{3/4} \cdot (4-k)^{-1/4} \\
 C_G & \equiv & (p-1) \cdot \left( C_{G0} C_{Gk}^k C_{Gkk}^{k^2} C_{Gp}^p C_{Gpk}^{pk} C_{Gpkk}^{pk^2} C_{Gpp}^{p^2} C_{Gppk}^{p^2k} C_{Gppkk}^{p^2k^2} \right)^{1/(4-k)} \cdot \frac{1}{3-k} \cdot \left( \frac{p-2}{p-1} \right)^{p-1} \nonumber \\
  & & \cdot (17-4k)^{\frac{-kp-5k+4p+12}{4(4-k)}} \cdot (4-k)^{\frac{3kp-5k-12p+12}{4(4-k)}} \cdot \frac{\Gamma(\frac{5}{4}+\frac{p}{4}) \Gamma ( \frac{p}{4}+\frac{19}{12} ) \Gamma ( \frac{p}{4}-\frac{1}{12} )}{ \Gamma(\frac{7}{4}+\frac{p}{4})(p+1)} \\
 C_H & \equiv & (p-1) \cdot \left( C_{H0} C_{Hk}^k C_{Hkk}^{k^2} C_{Hp}^p C_{Hpk}^{pk} C_{Hkk}^{pk^2} C_{Hpp}^{p^2} C_{Hppk}^{p^2k} C_{Hppkk}^{p^2k^2} \right)^{1/(4-k)} \cdot \frac{1}{3-k} \cdot \left( \frac{p-2}{p-1} \right)^{p-1} \nonumber \\
 & & \cdot (17-4k)^{\frac{p+2}{4}} \cdot (4-k)^{\frac{2-3p}{4}} \cdot \frac{\Gamma(\frac{5}{4}+\frac{p}{4}) \Gamma ( \frac{p}{4}+\frac{19}{12} ) \Gamma ( \frac{p}{4}-\frac{1}{12} )}{ \Gamma(\frac{7}{4}+\frac{p}{4})(p+1)} 
\end{eqnarray}

\IfFileExists{\jobname.bbl}{}
 {\typeout{}
  \typeout{******************************************}
  \typeout{** Please run "bibtex \jobname" to optain}
  \typeout{** the bibliography and then re-run LaTeX}
  \typeout{** twice to fix the references!}
  \typeout{******************************************}
  \typeout{}
 }


\end{document}